\documentclass[namedreferences,final]{SolarPhysics}

\usepackage{graphicx}
\usepackage{url}
\usepackage[optionalrh]{spr-sola-addons}
\usepackage{rotating}

\begin{document}
\begin{article}
\begin{opening}


\title{Two-Dimensional Spectroscopy of Photospheric Shear Flows in a Small
    $\delta$ Spot}

\author{C.~\surname{Denker}$^{1,2}$\sep
        N.~\surname{Deng}$^{2,3}$\sep
        A.~\surname{Tritschler}$^{4}$\sep
        V.~\surname{Yurchyshyn}$^{2,5}$}

\runningauthor{Denker et al.}
\runningtitle{Two-Dimensional Spectroscopy of Photospheric Shear Flows}

\institute{$^{1}$ Astrophysikalisches Institut Potsdam,\\
                  An der Sternwarte 16,
                  D-14482 Potsdam,
                  Germany,
                  email: \url{cdenker@aip.de}\\
           $^{2}$ New Jersey Institute of Technology,
                  Center for Solar-Terrestrial Research,\\
                  323 Martin Luther King Boulevard,
                  Newark, NJ~07102, U.S.A.\\
           $^{3}$ California State University Northridge,
                  Department of Physics and Astronomy,\\
                  18111 Nordhoff St,
                  Northridge, CA 91330, U.S.A.,
                  email: \url{na.deng@csun.edu}\\
           $^{4}$ National Solar Observatory/Sacramento Peak,\\
                  P.O.~Box 62,
                  Sunspot, NM~88349, U.S.A.,
                  email: \url{ali@nso.edu}\\
           $^{5}$ Big Bear Solar Observatory,\\
                  40386 North Shore Lane,
                  Big Bear City, CA 92314, U.S.A.,
                  email: \url{vayur@bbso.njit.edu}}

\date{Submitted: \ldots / Accepted: \ldots / Published online: \ldots}

%
%

\begin{abstract}
In recent high-resolution observations of complex active regions, long-lasting
and well-defined regions of strong flows were identified in major flares and
associated with bright kernels of visible, near-infrared, and X-ray radiation.
These flows, which occurred in the proximity of the magnetic neutral line,
significantly contributed to the generation of magnetic shear. Signatures of
these shear flows are strongly curved penumbral filaments, which are almost
tangential to sunspot umbrae rather than exhibiting the typical radial
filamentary structure. Solar active region NOAA~10756 was a moderately complex,
$\beta\delta$ sunspot group, which provided an opportunity to extend previous
studies of such shear flows to quieter settings. We conclude that shear flows
are a common phenomenon in complex active regions and $\delta$ spots. However,
they are not necessarily a prerequisite condition for flaring. Indeed, in the
present observations, the photospheric shear flows along the magnetic neutral
line are not related to any change of the local magnetic shear. We present
high-resolution observations of NOAA~10756 obtained with the 65-cm vacuum
reflector at {\it Big Bear Solar Observatory} (BBSO). Time series of
speckle-reconstructed white-light images and two-dimensional spectroscopic data
were combined to study the temporal evolution of the three-dimensional vector
flow field in the $\beta\delta$ sunspot group. An hour-long data set of
consistent high quality was obtained, which had a cadence of better than
30~seconds and sub-arcsecond spatial resolution.
\end{abstract}


\end{opening}

%
%

\section{Introduction}

The solar surface is highly dynamic and magnetic fields on various spatial
scales, from active regions and sunspots to network fields and faculae, are
intricately intertwined with the plasma motions. Photospheric flow fields are
important in the context of solar eruptive phenomena such as flares,
filament/prominence eruptions, and coronal mass ejections (CMEs). In
high-resolution studies of solar active region NOAA~10486, \inlinecite{Yang2004}
and \inlinecite{Deng2006} find strong shear flows close to the magnetic neutral
line, which are attributed to the slow energy build-up before its sudden release
in the flare. The local magnetic fields were highly non-potential and in this
case magnetic shear was the source of the free energy. However, the question
remains open if magnetic shear already existed when the magnetic field emerged.
This is of particular interest at the interface of solar dynamo theory and the
theory of flares and CMEs, since magnetic shear is related to highly-twisted
magnetic flux tubes, which can be generated by kink instability while the flux
tube rises from the bottom of the solar convection zone \cite{Fisher2000}.

\inlinecite{Amari2000} showed that twisted flux tubes play a crucial role in the
theory of large-scale solar eruptive phenomena such as CMEs and two-ribbon
flares. If photospheric twist is applied to the footpoints of an arcade-like
magnetic field structure, the filament might become unstable and erupt. In this
scenario, $\delta$ spots play a distinct role as the origin of a complex
magnetic topology in the corona, which makes it conducive to eruptions.
Therefore, multi-polar fields are important in the study and modeling of flares,
filament/prominence eruptions, and CMEs. The evolution of flares can be
explained by a magnetic breakout process first described by
\inlinecite{Antiochos1999}, {\it i.e.}, magnetic reconnection at a null point
that is located high in the corona is opening initially low-lying sheared fields
leading to the flare, filament eruption and CME. \inlinecite{Aulanier2000}
further elaborate that a three-dimensional magnetic null point exists in the
corona above the $\delta$ spot and that reconnection occurs at the null point
long before the flare. However, there is an on-going debate centered on the
exact location and type of reconnection. For example, \inlinecite{Moore2001}
present an opposing view in terms of the standard bipolar model for eruptive
flares, where they conceptually tie the sheared and twisted core fields of a
bipolar active region to a runaway ``tether-cutting'' via impulsive
reconnection.

In recent high-resolution observations, many phenomena were discovered, which
are directly related to eruptive phenomena such as the rapid penumbral decay as
a consequence of flares (\opencite{Wang2004}; \opencite{Deng2005};
\opencite{Liu2005}; \opencite{Sudol2005}), the detection of photospheric shear
flows near the magnetic neutral line and close to flare kernels
(\opencite{Yang2004}; \opencite{Deng2006}), and the first near-infrared (NIR)
continuum observations of flares \cite{Xu2004b}. Flare-associated shear flows
are  not a new phenomenon. During the major flares in August 1972,
\inlinecite{Zirin1973} measured spectroscopically strong photospheric velocity
discontinuities of up to 6~km~s$^{-1}$ in the vicinity of neutral lines. In
more recent observations, converging Doppler velocities were observed in the
vicinity of $\delta$ spots. ({\it e.g.}, \inlinecite{Lites2002}). These flows
were interpreted as an interleaved system of field lines, where the outward
Evershed flow sharply bends downward and returns to the solar interior.

In this study, we use two-dimensional spectroscopy to gain access to Doppler
velocities. Complementary observations of horizontal flows derived with local
correlation tracking (LCT, \opencite{November1988}) allow us to obtain a more
complete picture of the intricate flow field related to $\delta$ spots. These
high-cadence observations (30~seconds or less) of the photosphere in
combination with high-resolution observations of transition region and corona
have the potential to provide a comprehensive picture of rapid changes related
to magneto-convection and magnetic field evolution.

In Section~2, we introduce the observations, provide detailed accounts of the
data calibration, and discuss their limitations in the context of our scientific
objectives. The results are presented in Section~3, where we describe the
intricate, three-dimensional flow fields encountered in NOAA~10756 with a
particular emphasis on the $\delta$ configuration. We discuss the results in
Section~4 and conclude this study of complex flow fields in sunspots with a
summary of the most important results in Section~5.

%
%

\section{Observations}

In the Summer of 2005, we obtained the first visible-light data with the 65-cm
vacuum reflector and its new high-order adaptive optics (AO) system.
Two-dimensional imaging spectroscopy and speckle masking imaging were employed
to measure, besides other quantities, solar vector flow fields.

NOAA~10756 was selected as a promising target to study the dynamics in an active
region prior to an eruptive event, since it showed some moderate magnetic
complexity including a small $\delta$ configuration in its southern part.
However, the region did not produce any major M- or X-class flares during its
disk passage. In principle, scenarios with small flares (and even without any
flares) should receive the same attention as X-class events or geo-effective
CMEs. Otherwise, our views of such events might be skewed and we associate
phenomena with eruptive events that might be common to non-eruptive scenarios as
well. This is particularly true for high-resolution observations, which are rich
in detail and highly dynamic.

At the beginning of its disk passage on 26~April 2005, NOAA~10756 produced
several C-class flares. The largest event was an impulsive C5 flare. On
27~April, the region grew rapidly, predominantly in a $\delta$ spot attached to
the southern portion of the large sunspot. Despite its growth, NOAA~10756 only
produced lower-level C-class events. From 29~April to 1~May, the active region
reached its maximum area (about 1000 millionths of a solar hemisphere) before
it started to decay. We observed the moderately-complex $\delta$ configuration
on 2~May 2005, when it was located at heliographic coordinates
S8$^{\circ}$~W23$^{\circ}$ ($\mu = \cos\theta \approx 0.92$). At this time the
region was already decaying, however, it was still classified as a
$\beta\delta$ region. Only during the last few days of its disk passage a C7
and two M1 flares occurred, while the active region continued to decay.

\begin{kaprotate}
\begin{table}
\caption{BBSO Observing Characteristics on 2 May 2005\label{TAB01}}
\begin{tabular}{lrllrrcll}
\hline
Type                             & $\lambda$             & Start       & End         & $\Delta t$  & $n$ & FOV                                            & $s$                         & $\delta t$\\
\hline
Speckle Reconstruction           & $600 \pm 5$~nm        & 17:01:07~UT & 17:30:57~UT & 30~s           &  60 & $ 80^{\prime\prime} \times  80^{\prime\prime}$ & $0.08^{\prime\prime}$ pixel$^{-1}$ & 10~ms \\
Speckle Reconstruction$^{\rm a}$ & $600 \pm 5$~nm        & 17:40:33~UT & 18:09:49~UT & 30~s           &  60 & $ 80^{\prime\prime} \times  80^{\prime\prime}$ & $0.08^{\prime\prime}$ pixel$^{-1}$ & 10~ms \\
VIM                              & Fe\,{\sc i} 630.15~nm & 16:59:25~UT & 17:28:26~UT & 16~s           & 105 & $ 80^{\prime\prime} \times  80^{\prime\prime}$ & $0.16^{\prime\prime}$ pixel$^{-1}$ & 80~ms \\
VIM                              & Fe\,{\sc i} 630.15~nm & 17:38:39~UT & 18:08:00~UT & 16~s           & 110 & $ 80^{\prime\prime} \times  80^{\prime\prime}$ & $0.16^{\prime\prime}$ pixel$^{-1}$ & 80~ms \\
DVMG                             & Ca\,{\sc i} 610.30~nm & 15:53:10~UT & 22:30:40~UT & $\approx 80$~s & 220 & $288^{\prime\prime} \times 288^{\prime\prime}$ & $0.56^{\prime\prime}$ pixel$^{-1}$ & 30~ms$^{\rm b}$ \\
Filtergram                       & H$\alpha$ 656.30~nm   & 15:54:01~UT & 22:31:31~UT & $\approx 80$~s & 248 & $300^{\prime\prime} \times 300^{\prime\prime}$ & $0.58^{\prime\prime}$ pixel$^{-1}$ & 60~ms \\
\hline
\multicolumn{9}{p{190mm}}{\rule{3ex}{0mm}NOTE.---The symbol $\lambda$ refers
    to the wavelength, $\Delta t$ to the cadence, $n$ to the number of
    observations in each time series, $s$ to the image scale, and $\delta t$ to
    the exposure time. The speckle reconstructions and VIM data were taken with
    the 65-cm vacuum reflector whereas the DVMG and filtergrams were obtained as
    context data with the 25-cm refractor.}\\
\multicolumn{9}{p{190mm}}{\rule{3ex}{0mm}$^{\rm a}$The ten-minute gap between
    the two high-resolution data sets occurred, since the optical trains had to
    be re-aligned to compensate the pupil wobble on the AO wavefront sensor.}\\
\multicolumn{9}{p{190mm}}{\rule{3ex}{0mm}$^{\rm b}$The effective exposure time
    of the magnetograms is actually larger ($\approx$ two~seconds), since 64 individual
    exposures are combined to determine the degree of polarization Stokes
    $V/I$.}
\end{tabular}
\end{table}
\end{kaprotate}

\begin{figure}[p]
\centerline{\includegraphics[height=179mm]{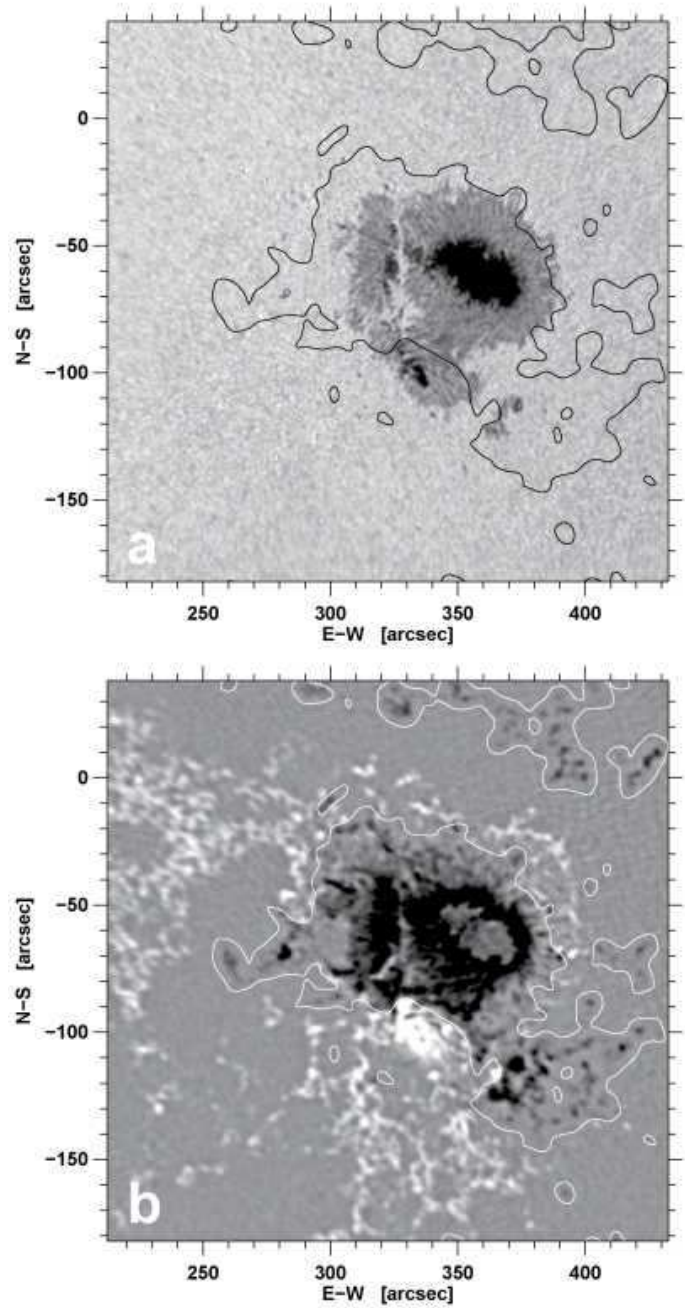}}
\caption{(a) Ca\,{\sc i} 610.3~nm line-wing filtergram and (b) longitudinal
    magnetogram obtained with BBSO's DVMG system at 17:52~UT on 2~May 2005. The
    magnetogram is displayed between $\pm 500$~G and the black (a) and white (b)
    contours represent the magnetic polarity inversion line.}
\label{FIG01}
\end{figure}

The observations of NOAA~10756 are mainly based on observations with BBSO's
65-cm vacuum reflector and 25-cm refractor. Table~\ref{TAB01} provides an
overview of the observing characteristics. Both telescopes have a suite of
instruments dedicated to active-region monitoring and space-weather research
\cite{Gallagher2002}. The observations with the 25-cm telescope provide the
context for the high-resolution observations. Two Apogee KX260 CCD cameras with
$512 \times 512$ pixel obtain filtergrams in the strong chromospheric
absorption lines Ca\,{\sc ii}\,K and H$\alpha$ with a one-minute cadence. The
image scale is about $0.58^{\prime\prime}$ pixel$^{-1}$ for both cameras
corresponding to a field-of-view (FOV) of about $300^{\prime\prime} \times
300^{\prime\prime}$. The third optical bench on the 25-cm telescope is reserved
for the {\it Digital Vector Magnetograph} (DVMG, \opencite{Spirock2001}). A
quasi-continuum image and a line-of-sight (LOS) magnetogram are shown in
Figure~\ref{FIG01}. We used an intensity image and magnetogram of the {\it
Michelson Doppler Imager} (MDI, \opencite{Scherrer1995}) to establish the exact
image scale and coordinates for the DVMG data. Before aligning MDI and DVMG
data the solar position angle has to be taken into account. The image scale of
the DVMG is $0.56^{\prime\prime}$ pixel$^{-1}$ corresponding to a FOV of
$288^{\prime\prime} \times 288^{\prime\prime}$. The DVMG FOV is
$220^{\prime\prime} \times 220^{\prime\prime}$ after solar $P$-angle
correction. The $x$- and $y$-axes in Figure~\ref{FIG01} are given in disk
center coordinates and can be used as a reference for the high-resolution
observations. We usually depict the high-resolution images without rotation
correction. The DVMG line-wing image is commonly used as a reference, if we
have to co-align the H$\alpha$ filtergrams, the speckle reconstructions, and
two-dimensional spectroscopic data. Note that the line-wing images are used as
a substitute for continuum images.

The high-resolution instrumentation of the 65-cm telescope is located in a
coud\'e laboratory. It consists of the high-order AO system, two imaging
spectro-polarimeters for observations in the visible and near infrared
wavelength regions (\opencite{Denker2003a}, \citeyear{Denker2003b}), and a fast
CCD camera system for image restoration \cite{Denker2005b}. The seeing
characteristics at BBSO were evaluated as part of the site survey for the {\it
Advanced Technology Solar Telescope} (ATST) and the site-specific results are
summarized in \inlinecite{Verdoni2007}. The high-order AO system
(\opencite{Rimmele2004a}; \opencite{Rimmele2004a}) is a collaboration between
BBSO and the {\it National Solar Observatory} (NSO). The optical set-up and
performance of the BBSO system are presented in \inlinecite{Denker2007b}. One
limitation of the AO system should be noted here, since it affects the data
collection in this study. A misalignment of the declination and right ascension
axes led to a wobble of the pupil image on the wavefront sensor. As a result,
the light path of the AO system and the post-focus instrumentation had to be
realigned every 30~minutes. This leads to a data gap of about five to ten
minutes between the 30-minute observing sequences.

\begin{figure}[t]
\centerline{\includegraphics[width=80mm]{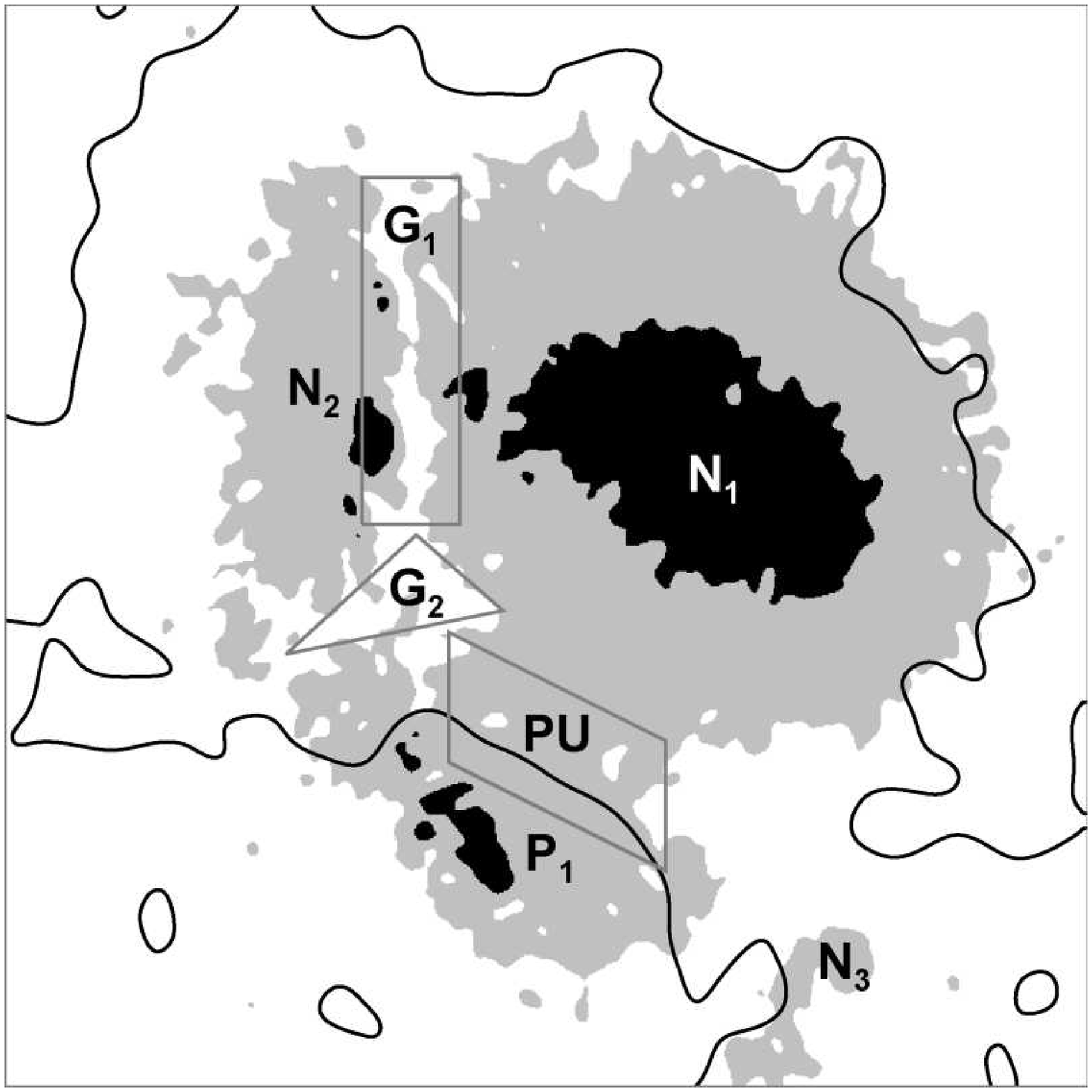}}
\caption{Schematic sketch of the small $\delta$ configuration in NOAA~10756.
    Granulation, penumbrae, and umbrae are indicated as white, gray, and black
    regions, respectively. The black line represents the magnetic neutral line.
    The labels correspond to features discussed in the text.}
\label{FIG02}
\end{figure}

Continuum images at $600 \pm 5$~nm of the North-East and South-East parts of
the active region were obtained from 17:01~UT to 17:31~UT and 17:40~UT to
18:10~UT, respectively. The time series of speckle restored images has a
cadence of 30~seconds, {\it i.e.}, we restored a total of $2 \times 60$ images.
Each restored image is based on 100 short-exposure (ten~ms) images, which were
selected from 200 frames captured at 15~frames~s$^{-1}$. The FOV was about
$79.5^{\prime\prime} \times 79.5^{\prime\prime}$ with an image scale of
$0.078^{\prime\prime}$ pixel$^{-1}$. A target was inserted in a focal plane
following the tip-tilt mirror to accurately align the channels for image
restoration and  two-dimensional spectroscopy. The image restoration procedure
has been described in detail in \inlinecite{Denker2005b}.

In this study, we present for the first time scientific results obtained with a
new two-dimensional imaging spectrometer. In 2007, we will add polarization
optics to the spectrometer, which will enable us to measure the full Stokes
vector. The design of the imaging magnetograph systems at BBSO has been
outlined in \inlinecite{Denker2003a} and \inlinecite{Denker2003b}. As for now,
we use {\it Visible-light Imaging Magnetograph} (VIM) just as a spectrometer.
The imaging spectrometer was carefully tested, which includes a detailed
characterization of the Fabry-P\'erot etalon \cite{Denker2005a}. The
two-dimensional spectroscopic observations in the Fe\,{\sc i} 630.15~nm line
basically follow the continuum observations. However, they have a slightly
larger FOV of $84.4^{\prime\prime} \times 84.4^{\prime\prime}$. The CCD camera
was operated in a two $\times$ two pixel binning mode. Thus, the image scale
increased to $0.165^{\prime\prime}$ pixel$^{-1}$ compared to the continuum
observations. The cadence of the spectral line scans was 16~seconds and a total
of 214 scans were obtained. The filtergrams were taken at 91 wavelength points
in the Fe\,{\sc i} line separated by 1.2~pm corresponding to a scanned
wavelength interval of 0.11~nm. The exposure time of the filtergrams was 80~ms.

%
%

\section{Results}

Figures~\ref{FIG01} and \ref{FIG02} depicts the leading sunspot in active
region NOAA 10756 with predominantly negative polarity ({\it N}$_1$ and {\it
N}$_2$). The labels printed in italics refer to the schematic sketch shown in
Figure~\ref{FIG02}, which should serve as a guide to the many (small-scale)
features discussed in the following sections. Only a small $\delta$
configuration with opposite polarity ({\it P}$_1$) can be found in the southern
part of the region. The trailing positive polarity plage extends further to the
East -- well beyond the FOV shown in Figure~\ref{FIG01}. Flux concentrations
are limited to small pores or magnetic knots, which are invisible in MDI
magnetograms but sometimes recognizable in DVMG magnetograms at higher spatial
resolution. The leading part of NOAA~10756 consists of three distinct parts: a
dominant elliptical spot of negative polarity ({\it N}$_1$), a linear
arrangement (oriented along the North-South axis) of small sunspots with a
rudimentary penumbra ({\it N}$_2$) of the same polarity as the main sunspot,
and a small $\delta$ configuration ({\it P}$_1$). The North-South aligned spots
({\it N}$_2$) are separated by a narrow granular region ({\it G}$_1$, vertical
box in Figure~\ref{FIG02}) from the main spot, while the penumbrae ({\it PU},
diagonal box in Figure~\ref{FIG02}) of the $\delta$ and main spot are fused
together. The magnetic polarity inversion line surrounding the negative flux of
the active region was derived from a low-pass filtered version of the
magnetogram. It is superimposed as black and white contours on the Ca\,{\sc
ii}\,K line-wing image and magnetogram shown in Figure~\ref{FIG01}. This
neutral line is also depicted in the high-resolution images.

\begin{figure}[p]
\centerline{\includegraphics[height=164mm]{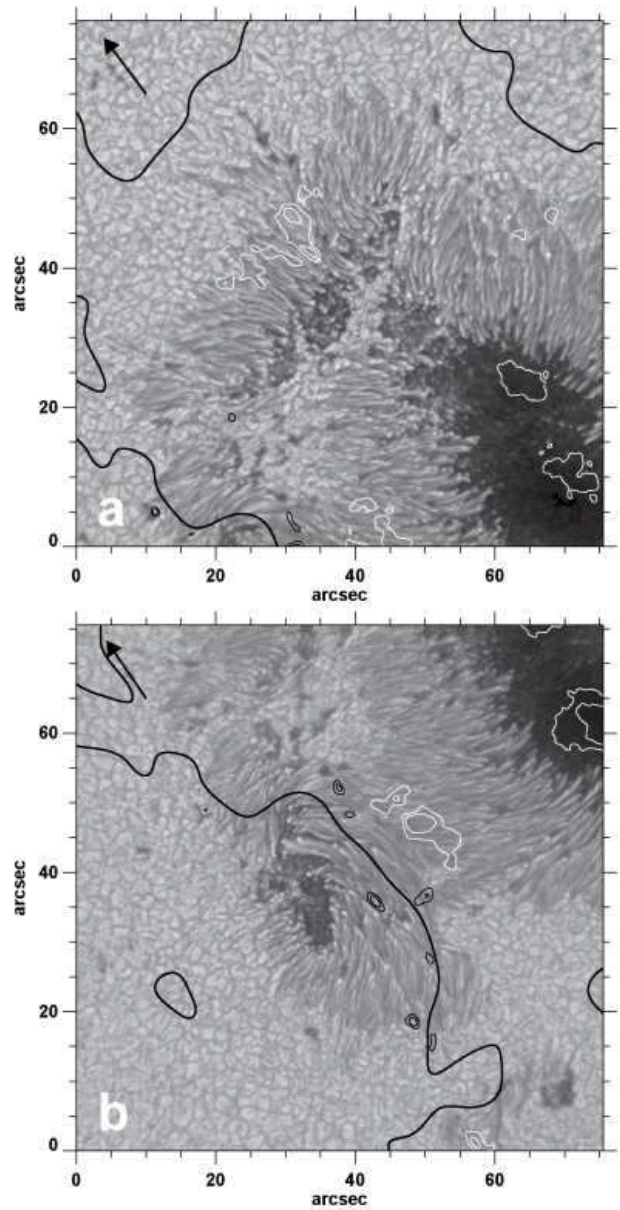}}
\caption{Speckle reconstructions of (a) the North-East (17:09~UT) and (b)
    South-East (17:55~UT) parts of solar active region NOAA~10756 obtained on
    2~May 2005 with BBSO's 65-cm vacuum reflector. The thick black contours
    represent the magnetic polarity inversion line. The thin black (white)
    contours outline regions with LOS speeds of 1.0 and 1.2~km~s$^{-1}$ ($-1.0$
    and $-0.8$~km~s$^{-1}$), respectively (see Figure~\ref{FIG04}). The black
    arrows in the upper right corners indicate the direction toward disk
    center. The $\delta$ spot is located in the center of the lower panel.}
\label{FIG03}
\end{figure}

Since the FOV of the high-resolution observations is only $80^{\prime\prime}
\times 80^{\prime\prime}$, we decided to obtain two different data sets zooming
in on several small sunspots to the East ({\it N}$_2$) and a small $\delta$
configuration ({\it P}$_1$) to the south of the main sunspot ({\it N}$_1$).
Several small light-bridges branch out from the about $5^{\prime\prime}$-wide
channel of granulation ({\it G}$_1$), which separate individual umbral cores.
All umbral cores contain numerous umbral dots, which consist of peripheral
umbral dots with histories related to the migration of penumbral grains and a
background of fuzzier umbral dots. This linear array of sunspots formed a
well-organized rudimentary penumbra on its eastern side. The $\delta$ spot
({\it P}$_1$) is located to the South-East of these small spots ({\it N}$_2$),
separated by a triangular shaped region of granulation. The main sunspot
connects to the remaining side of the triangle. Several tiny pores or magnetic
knots permeate this region of granulation ({\it G}$_2$, triangular box in
Figure~\ref{FIG02}), which is visible in both panels of Figure~\ref{FIG03}. The
$\delta$ spot consists of a single umbra containing several umbral dots, which
are, however, much fainter and less densely packed compared to the
aforementioned small sunspots. The umbral core is about $15^{\prime\prime}
\times 5^{\prime\prime}$. The penumbral filaments, which face away from the
main spot, point radially outward, whereas the filaments closest to the main
spot are strongly curved and wrapped around the umbral core of the $\delta$
spot. These almost-tangential filaments are the first visual indications of
highly twisted magnetic field lines.

Tracing these individual penumbral filaments and penumbral grains in time
series of speckle reconstructions reveals a counter-clockwise twist. The
penumbral filaments of the main spot in the vicinity of the $\delta$ spot show
exactly the same twist leading to strong shear flows at the interface of the
two penumbrae. The penumbral segments, where penumbral filaments and grains
follow curved trajectories, are much wider than their counterparts with normal
radial motions. This might be a direct result of the Evershed flow. The
proximity of two penumbrae ({\it PU}) with colliding outflows leaves only two
possibilities: the subduction of one flow pattern or the horizontal deflection
of the opposing flows. Our observations of penumbral filaments, which are
tangential to the umbra of the $\delta$ spot well support mainly the latter
case. Strong shear flows are thus created, which will be aligned with the
polarity inversion line, if the two sunspots are of opposite polarity as in our
case. However, the subduction scenario put forward in \inlinecite{Lites2002}
would still apply to the locations where the dark penumbral filaments carrying
the Evershed flows terminate, {\it i.e.}, bend downward and return their
respective flows to the solar interior. Strands of penumbral filaments extend
for more than $25^{\prime\prime}$ while winding around the $\delta$ spot. These
strands show indications of a braided structure, which could be interpreted as
locally-confined subduction or horizontal twists in the flow channels
associated with the Evershed effect. Finally, the whirlpool-like magnetic
structure ({\it N}$_3$) near the western tip of the $\delta$ spot ({\it P}$_1$)
can serve as an example that localized shear flows and twist are present in
this active region even on scales of about $10^{\prime\prime}$.

\begin{figure}[p]
\centerline{\includegraphics[height=164mm]{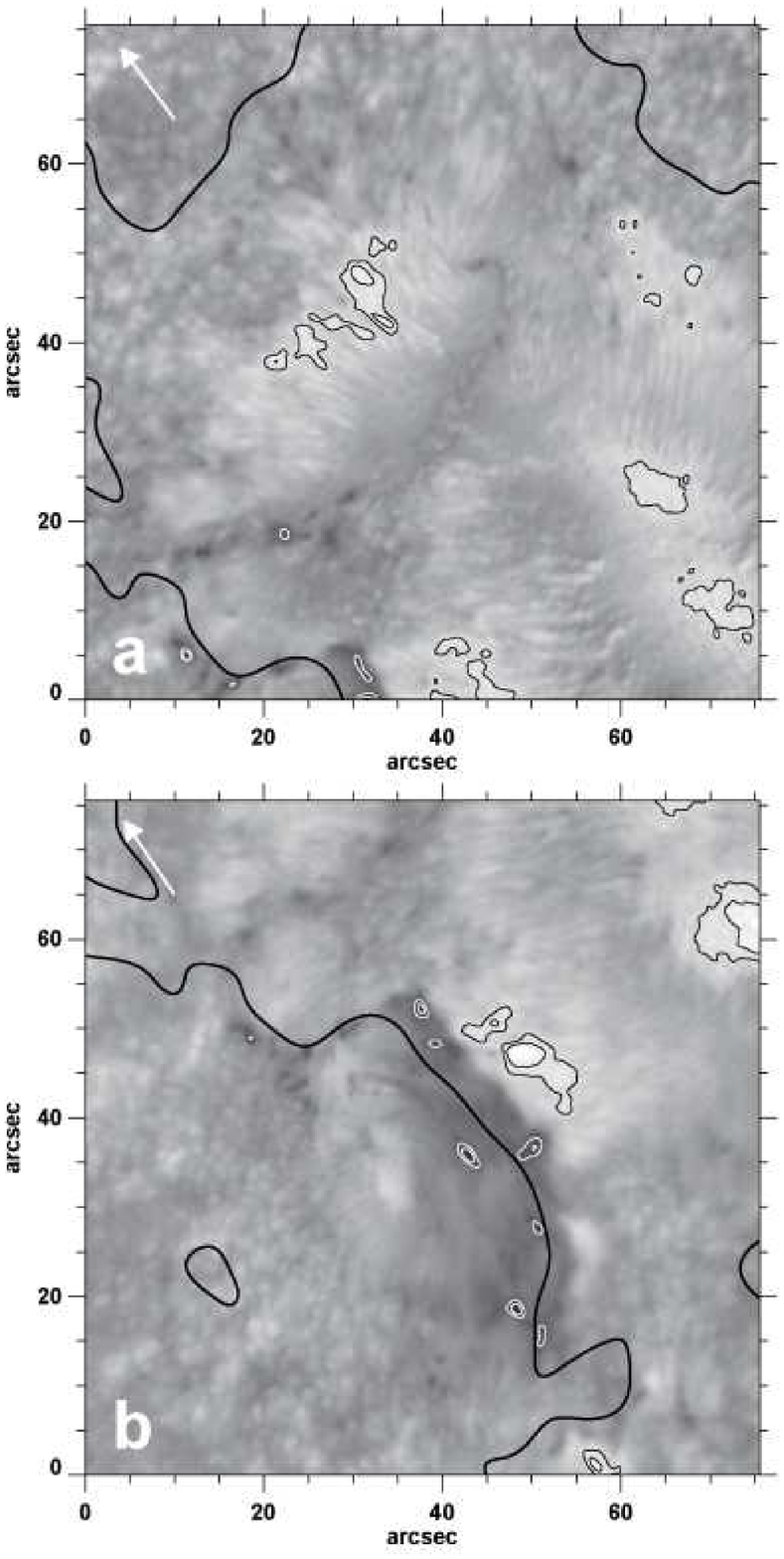}}
\caption{Dopplergrams displayed in the range of $\pm 1.5$~km~s$^{-1}$ of (a) the
    North-East (17:09~UT) and (b) South-East (17:55~UT) parts of solar active
    region NOAA~10756 obtained on 2~May 2005 with BBSO's 65-cm vacuum reflector.
    The thick black contours represent the magnetic polarity inversion line.
    The thin white (black) contours outline regions with LOS speeds of 1.0 and
    1.2~km~s$^{-1}$ ($-1.0$ and $-0.8$~km~s$^{-1}$), respectively. The white
    arrows in the upper right corners indicate the direction toward disk center.
    The dopplergrams were slightly cropped to fit the FOV of the continuum
    speckle reconstructions.}
\label{FIG04}
\end{figure}

In Figure~\ref{FIG04}, we show dopplergrams obtained with the imaging
spectrometer. The Fe\,{\sc i} 630.15~nm line is broadened by the Zeeman effect.
This limitation should be kept in mind, when interpreting the findings based on
VIM data. The LOS velocity is determined by a Fourier-phase method
\cite{Schmidt1999}, which uses the entire line profile and is very insensitive
to noise. We follow the convention that redshifts are negative and blueshifts
are positive. Thus, dark areas in the dopplergrams shown in Figure~\ref{FIG04}
move away from the observer, while bright areas indicate motion towards the
observer. In the absence of an absolute wavelength calibration, we used quiet
Sun areas to establish a velocity calibration taking into account a convective
blueshift of 0.195~km~s$^{-1}$ \cite{Balthasar1988} for the Fe\,{\sc i}
630.15~nm line at $\mu = \cos\theta = 0.92$. Since we are mostly interested in
the morphology and evolution of the flow fields, a rough interpretation of LOS
velocities as up- and downflows can be justified. In the following, we
concentrate on the major features of the LOS flows. A summary of the typical
flow velocities encountered in and around the active region are presented in
Table~\ref{TAB02}.

\begin{table}[t]
\caption{Typical Velocities in Active Region NOAA~10756}
\begin{tabular}{lccccc}
\hline
Technique & Region & Quiet Sun       & Penumbra        & Umbra           & Flow Kernel\\
\hline
LOS       & 1      & $0.23 \pm 0.17$ & $0.44 \pm 0.27$ & $0.55 \pm 0.30$ & $\cdots$\\
LOS       & 2      & $0.23 \pm 0.17$ & $0.39 \pm 0.26$ & $0.64 \pm 0.30$ & $0.96 \pm 0.15$\\
LCT       & 1      & $0.46 \pm 0.27$ & $0.38 \pm 0.26$ & $0.26 \pm 0.14$ & $\cdots$\\
LCT       & 2      & $0.49 \pm 0.28$ & $0.35 \pm 0.27$ & $0.28 \pm 0.16$ & $0.90 \pm 0.39$\\
\hline
\multicolumn{6}{p{100mm}}{\rule{3ex}{0mm}NOTE.---The region indices identify the
    FOV in (1) the top panels and (2) the bottom panels shown in
    Figures~\ref{FIG03} and \ref{FIG04}. LCT and LOS refer to the magnitude of
    the velocities as measured with local correlation tracking and
    two-dimensional spectroscopy. All velocities are given in units of
    km~s$^{-1}$. The standard deviation of the velocities reflects the width of
    the frequency distribution for the various solar regions and is not to be
    confused with a measurement error.}
\end{tabular}
\label{TAB02}
\end{table}

Some of the strongest LOS downflows are concentrated in the small pores and
magnetic knots contained in the triangle of granulation ({\it G}$_2$)
connecting the main spot ({\it N}$_1$), the linear group of sunspots ({\it
N}$_2$), and the $\delta$ spot ({\it P}$_1$). Typical downflow velocities in
these small-scale flux concentrations range from 0.6 to 1.0~km~s$^{-1}$. The
rudimentary penumbra belonging to the linear group of spots ({\it N}$_2$) has
the same velocity sign as the center-side penumbra of the main spot ({\it
N}$_1$).

\begin{figure}[p]
\centerline{\includegraphics[height=172mm]{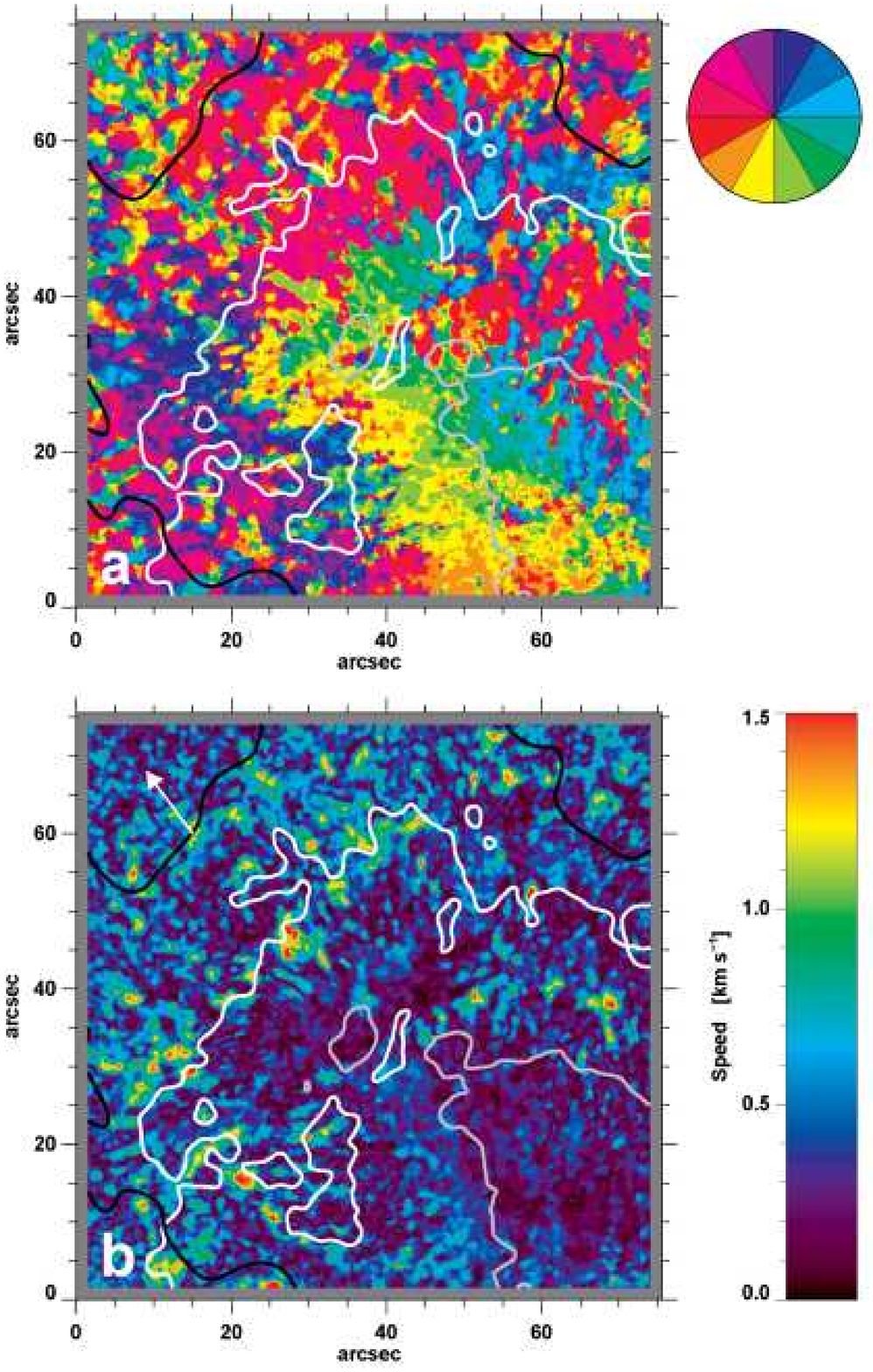}}
\caption{Horizontal flow fields in NOAA~10756 on 2~May 2005. LCT was applied to
    sequences of 60 speckle reconstructions obtained during the time period from
    17:01~UT to 17:31~UT (northern part). (a) The color-coded direction of the
    optical flows (see the ``compass rose'' in the upper right corner) and (b)
    the corresponding horizontal flow speeds. The white arrow indicates the
    direction toward disk center. The FOV is slightly smaller compared to
    Figure~\ref{FIG03} due to image rotation in the coud\'e laboratory.}
\label{FIG05}
\end{figure}

\begin{figure}[p]
\centerline{\includegraphics[height=172mm]{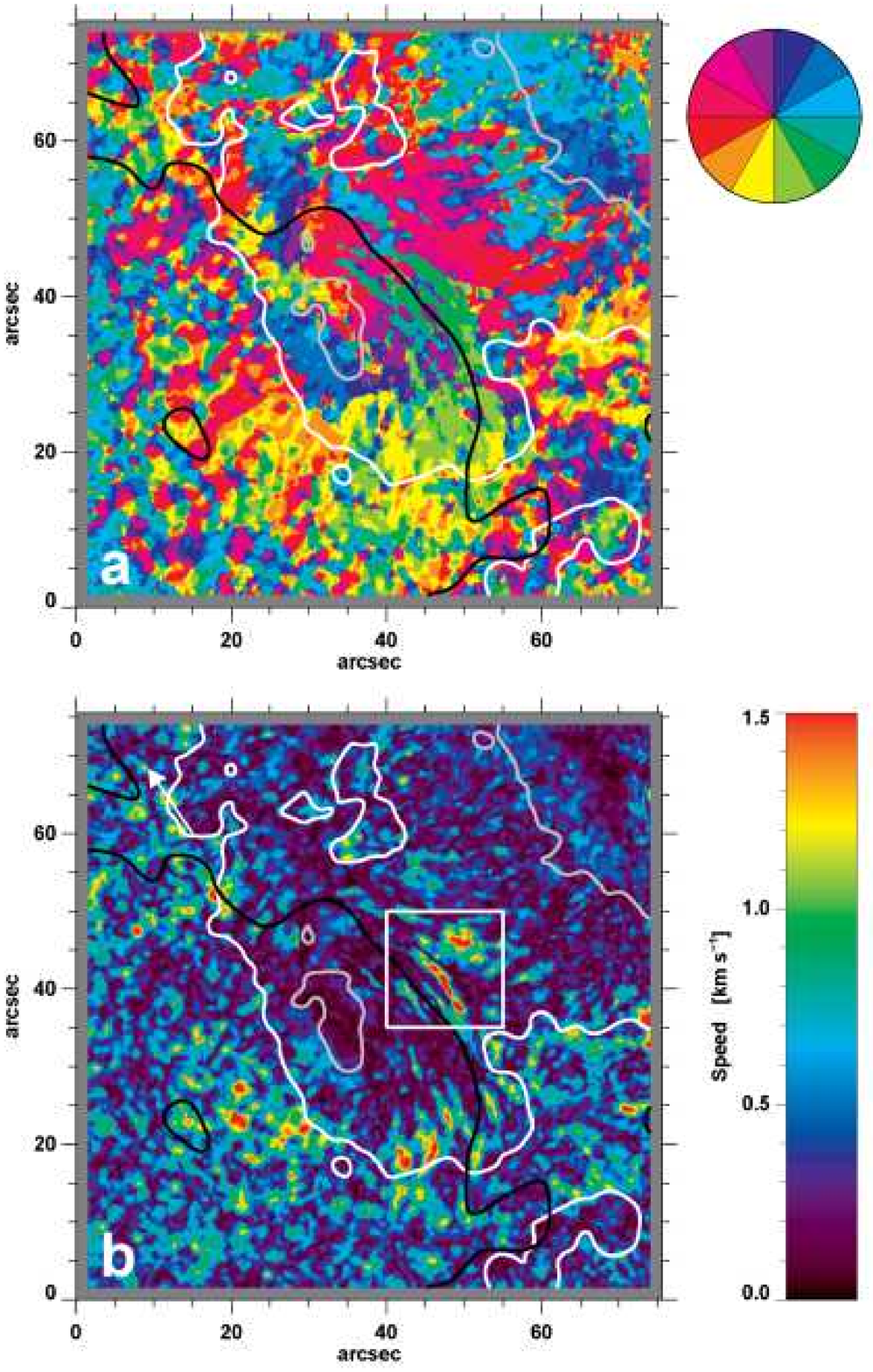}}
\caption{Horizontal flow fields in NOAA~10756 on 2~May 2005. LCT was applied to
    sequences of 60 speckle reconstructions obtained during the time period from
    17:40~UT to 18:10~UT (southern part). (a) The color-coded direction of the
    optical flows (see the ``compass rose'' in the upper right corner) and (b)
    the corresponding horizontal flow speeds. The white arrow indicates the
    direction toward disk center. The white box indicates the region with
    strong horizontal shear flows.}
\label{FIG06}
\end{figure}

The most conspicuous LOS velocity feature of the $\delta$ spot is its penumbra.
The line separating upflows and downflows (contained in the diagonal box {\it
PU} in Figure~\ref{FIG02}) basically follows the magnetic polarity inversion
line. However, it is displaced by about $5^{\prime\prime}$. A closer inspection
of the magnetograms reveals that the positive polarity penumbra of the $\delta$
spot contains several elongated inclusions of of small-scale negative polarity
features, {\it i.e.}, without smoothing several localized magnetic neutral
lines would exist. The neutral line shown in Figures~\ref{FIG01} to \ref{FIG06}
was derived from a smoothed magnetogram, which eliminate many of these small
``polarity islands'' and thus effectively decreases the majority flux
surrounding these locations. As a result, the magnetic neutral line is shifted
towards the $\delta$ spot. However, the separation line between up- and
downflows coincides well with the strongest magnetic field gradient. These
observations support the scenario in which the penumbrae of the $\delta$ spot
and the main spot behave primarily as separate entities, {\it i.e.}, the
Evershed flow is deflected. However, intrusions of opposite polarity flux might
occur near this interface, which appear as the aforementioned braided structure
of penumbral filaments. Localized subduction as suggested by
\inlinecite{Lites2002} might play a role in these cases. Only high-resolution
two-dimensional spectro-polarimetry with sub-arcsecond resolution can answer
this question.

Since the penumbral filaments on the western side are almost tangential, they
behave essentially as if they belong to the limb-side penumbra. Strong flow
kernels are embedded in both colliding penumbrae. The size of these elongated
flow kernels is about  $5^{\prime\prime} \times 2^{\prime\prime}$ and they are
aligned with the local penumbral filaments. Flows in these kernels are
typically larger than 0.8~km~s$^{-1}$ and reach almost 1.8~km~s$^{-1}$. No dark
features in the speckle restored continuum images (see Figure~\ref{FIG03}) are
associated with these flow kernels. Analyzing the time series of dopplergrams,
we find that the flow kernels are slowly evolving, long-lived features (at
least 30~minutes) that do not partake in the penumbral proper motions seen in
time series of speckle restored continuum images. Since no magnetograms with a
spatial resolution comparable to that of the two-dimensional spectroscopic data
or the speckle restored images was available, the questions remains open, if
the flow kernels are associated with intrusion of opposite polarity flux into
the the penumbral shear flow areas.

In Figures~\ref{FIG05} and \ref{FIG06}, we show high-resolution flow maps based
on time series of speckle reconstructed images. The two figures depict the
detailed proper motions in the northern and southern parts of active region
NOAA 10756, respectively. Averaging 60 individual flow maps over a period of
about 30~minutes only leaves persistent flow patterns. The smallest reliable
features in the flow maps have a size of about $1^{\prime\prime}$. For this
reason, we opted to use a color representation to indicate flow direction and
magnitude rather than relying on drawing vector maps with inferior resolution.
A 12-sector RGB color wheel has been added as a visualization guide in the
upper-right corners of the azimuth maps. In addition, contour lines of the
granulation/penumbra boundary (white) and the penumbra/umbra (gray) were
superimposed on the flow maps to ease the interpretation of the highly detailed
maps. The areas with granulation are characterized by small-scale patches
(having the size of individual granular cells) with large variations of the
azimuth angle. In addition, in regions with small-scale magnetic fields, {\it
i.e.}, brightenings in the line core intensity maps, we find some of the
largest proper motions. Typical horizontal velocities related to Fe\,{\sc i}
line gaps frequently reach 1.25~km~s$^{-1}$, whereas velocities in areas of
normal granulation are typically around 0.75~km~s$^{-1}$. A comparison of the
typical LCT and LOS velocity values is given in Table~\ref{TAB02} for different
regions and features in the active region.

The proper motions in penumbral regions retain the elongated character of the
penumbral filaments. The largest velocities are generally found near the
granulation/penumbra boundary, {\it i.e.} they are encountered in the vicinity
of the white contour lines in Figure~\ref{FIG05}b. Conversely, the lowest
velocities are found in the umbral regions of the sunspots. However, this might
be related to the low signal-to-noise ratio in these dark regions, where only
the faint umbral dots contribute to the correlation signal measured with the
LCT technique. The high LOS velocities, which were associated with the small
pores and magnetic knots in the previously mentioned triangular granular region
({\it G}$_2$), have no counterparts in the LCT maps. The most prominent feature
in the horizontal flow maps is located in the center of Figure~\ref{FIG06}b as
indicated by a white box. Two flow kernels, one elongated and the other with
circular shape, are clearly visible. The diameter of the circular kernel is
about $5^{\prime\prime}$ and the other kernel is about $3^{\prime\prime}$ wide
and stretches out to about $12^{\prime\prime}$. The velocity in these flow
kernels approaches 2.5~km~s$^{-1}$. In addition, they are co-aligned with flow
kernels in the LOS velocity maps. The two flow kernels are within
$5^{\prime\prime}$ of each other and separated by a line tracing the strongest
magnetic field gradient, which parallels the magnetic polarity inversion line
about $5^{\prime\prime}$ to the North-West. Compared to previous observations
of active region NOAA~10486 (see \opencite{Yang2004}, especially Figures~2b and
2c), the flow interface is harder to trace in the azimuth and magnitude maps,
since it does not exactly follow the magnetic polarity inversion line. However,
the proximity of green and purple colors $5^{\prime\prime}$ to the North-East
of the magnetic neutral lines indicates flows in opposite direction. Thus,
strong shear flows are present in NOAA~10756 as well.

\begin{figure}[t]
\centerline{\includegraphics[width=120mm]{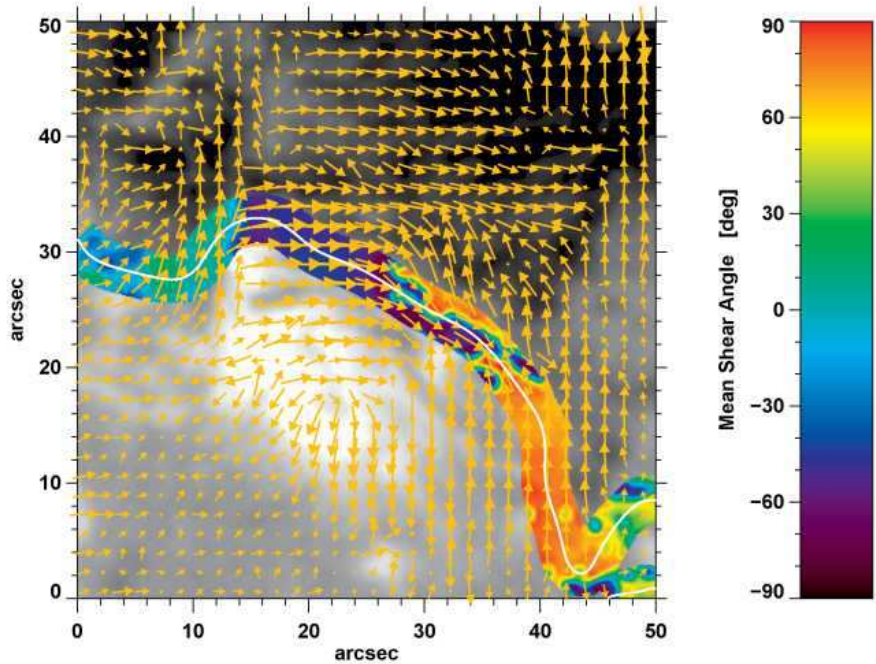}}
\caption{Vector magnetogram of the small $\delta$ configuration. The
    transverse magnetic field is given as orange vectors superposed on a
    gray-scale representation of the longitudinal magnetogram, which is scaled
    between $\pm 500$~G. The magnetic neutral line is shown in white. The
    magnetic shear angle is displayed in a neighborhood of $\pm 2^{\prime\prime}$
    along the magnetic neutral line. The magnetic shear angle is encoded
    according to the color index on the right.}
\label{FIG07}
\end{figure}

We define magnetic shear according to \inlinecite{Wang1994} as the angular
difference between the measured and calculated potential transverse field. In
Figure~\ref{FIG07}, we show in the background a longitudinal magnetogram of the
small $\delta$ configuration taken at about the same time as the data shown in
Figure~\ref{FIG01}. The FOV of the vector magnetogram is $50^{\prime\prime}
\times 50^{\prime\prime}$. The orange arrows indicate magnitude and direction
of the measured transverse field component. The $180^{\circ}$  ambiguity was
resolved by applying the acute angle method, which requires that the transverse
components of the observed and modeled (in this case potential) fields make an
cute angle. Recent studies in the framework of the SDO/HMI--CSAC Azimuth
Ambiguity Resolution Workshop \cite{Metcalf2006} showed that this method can
correctly resolve ambiguity for about 90\% of strong transverse magnetic
fields. The magnetic shear angle has been computed for a $2^{\prime\prime}$
neighborhood along both sides of the magnetic neutral line (shown as a thick
white line). The magnetic shear angle in the interval $[-90^{\circ},
+90^{\circ}]$ is mapped to a rainbow color-scale shown at the right side of
Figure~\ref{FIG07}. Two areas with distinct magnetic shear angles exist along
the magnetic neutral line. One area corresponds to the diagonal box {\it PU} in
Figure~\ref{FIG02}, where the photospheric shear flows are observed. The mean
magnetic shear angle is negative (blue) in this region. In contrast, the
magnetic shear angle along the magnetic neutral line immediately to the south
of is positive (red).

\begin{figure}[t]
\centerline{\includegraphics[width=100mm]{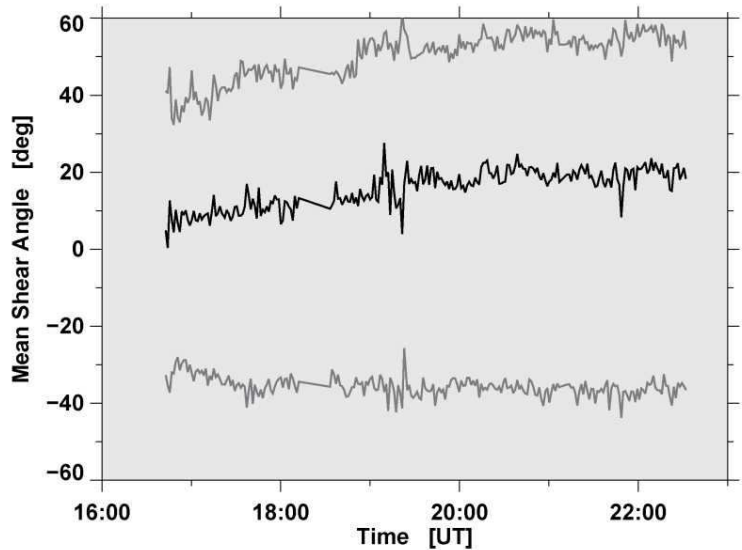}}
\caption{Temporal evolution of the mean magnetic shear angle along the
    entire magnetic neutral line shown in Figure~\ref{FIG07} (black curve). The
    temporal evolution of the mean magnetic shear angle for region {\it PU} with
    photospheric shear flows (negative shear) and for the region immediately to
    the South (positive shear) is given by the two gray curves at the bottom and
    top, respectively.}
\label{FIG08}
\end{figure}

Since we have 243 vector magnetograms during the time period from 16:43~UT to
22:32~UT, we can study the temporal evolution of the magnetic shear and relate
it to the observed photospheric shear flows. For this purpose, we defined a
neighborhood of $2^{\prime\prime}$ along both sides of the magnetic neutral
line as indicated in Figure~\ref{FIG07} but now for all vector magnetograms.
The average magnetic shear is computed for the entire neighborhood area of the
neutral line visible in the $50^{\prime\prime} \times 50^{\prime\prime}$ FOV of
Figure~\ref{FIG07}. Its temporal evolution is presented by the black curve in
Figure~\ref{FIG08}. The magnetic shear increased from about $+9^{\circ}$ at
17:00~UT to about $+21^{\circ}$ at 22:30~UT. However, this increase occurs
predominantly in the southern part of the neutral line with positive magnetoc
shear angle, whereas in proximity to  the region with photospheric shear flows,
the magnetic shear angle remains fairly constant at about $-36^{\circ}$. Shear
angle values around $\pm 30^{\circ}$ are also commonly encountered in
non-$\delta$ areas of sunspots. In $\delta$ spots with major flares, the
magnetic shear angle is often close to $\pm 90^{\circ}$ ({\it e.g.},
\opencite{Hagyard1984}).

\begin{figure}[t]
\centerline{\includegraphics[height=120mm]{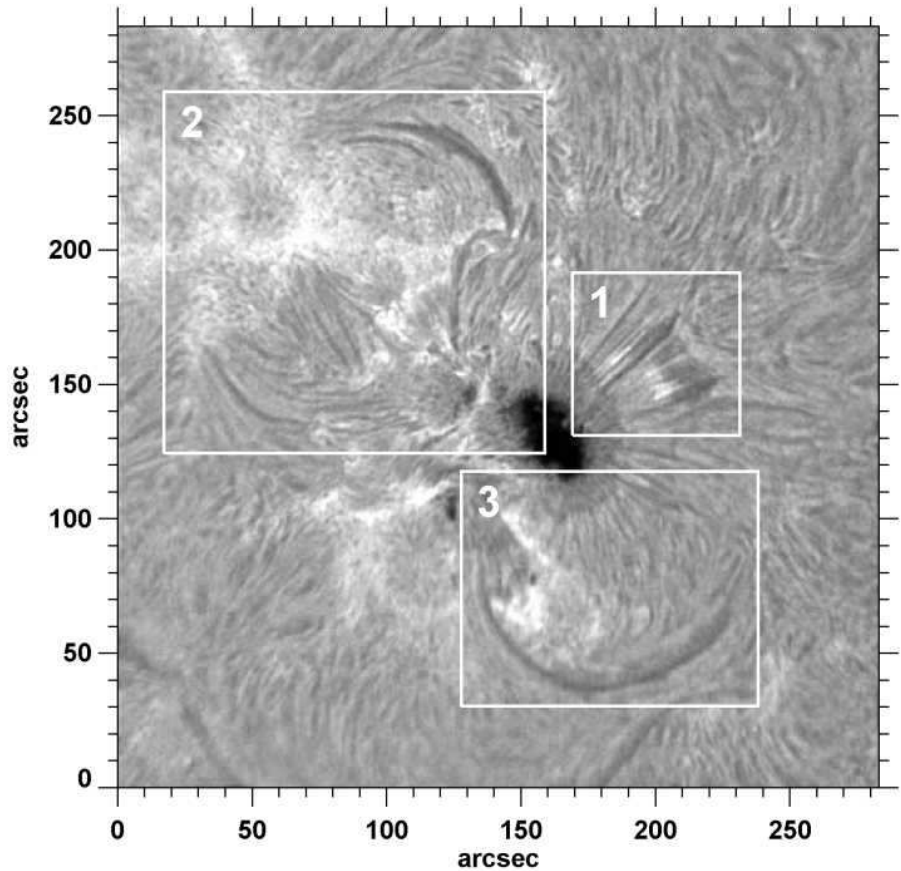}}
\caption{H$\alpha$ filtergram of NOAA~10756 obtained at 17:51~UT on 2~May 2005
    with BBSO's 25-cm refractor. The white boxes mark three H$\alpha$ loop
    systems discussed in the text.}
\label{FIG09}
\end{figure}

In addition to the photospheric observations, we obtained chromospheric data,
which provides some clues to the overall magnetic field topology. Three
distinct loop system indicated by white boxes in Figure~\ref{FIG09} can be
identified in the H$\alpha$ filtergrams: i) small loops connecting the main
spot with opposite polarity flux in the sunspot moat in the north-western part
of the active region, ii) large-scale loops associated with the linear group of
small sunspots ({\it N}$_2$) in the North-East, which terminate in a region of
extended plage, and iii) a large loop between the $\delta$ spot ({\it P}$_1$)
and the umbra of the main sunspot ({\it N}$_1$). All loop systems have in
common that they point away from the center of the main spot and leave the
central region of the sunspot group devoid of any H$\alpha$ loops. In
particular, no active-region filament formed along the polarity-inversion line.
In time series of H$\alpha$ filtergrams, we find some indications of activity
related to the loop systems, which is most prominent in the flux system towards
the North-East (Box~1 in Figure~\ref{FIG09}). Here, we find frequent
brightenings of the H$\alpha$ plage, which originates at the footpoint of a
small filament routed in the linear sunspot group. The $\delta$ spot filament
(Box~3) is fairly inactive with the exception of one short-term brightening
around 18:07~UT, which traveled at high speeds from the umbra of the main spot
to approximately the apex of the H$\alpha$ loop. Even though the connection of
the $\delta$ spot filament to the main sunspot might not be clear in
Figure~\ref{FIG09}, the trajectory of this brightening undoubtedly confirms
this connection. This feature is very similar to transient flow fields in
filaments, which were observed by \inlinecite{Chae2000} in high-resolution
times-series of H$\alpha$ filtergrams. In addition to these transient flows, we
find indications of twisted flows near the loop top. These twisted flows are
especially prominent during times when the absorption is increasing at the apex
of the loop.

\begin{figure}[t]
\centerline{\includegraphics[height=120mm]{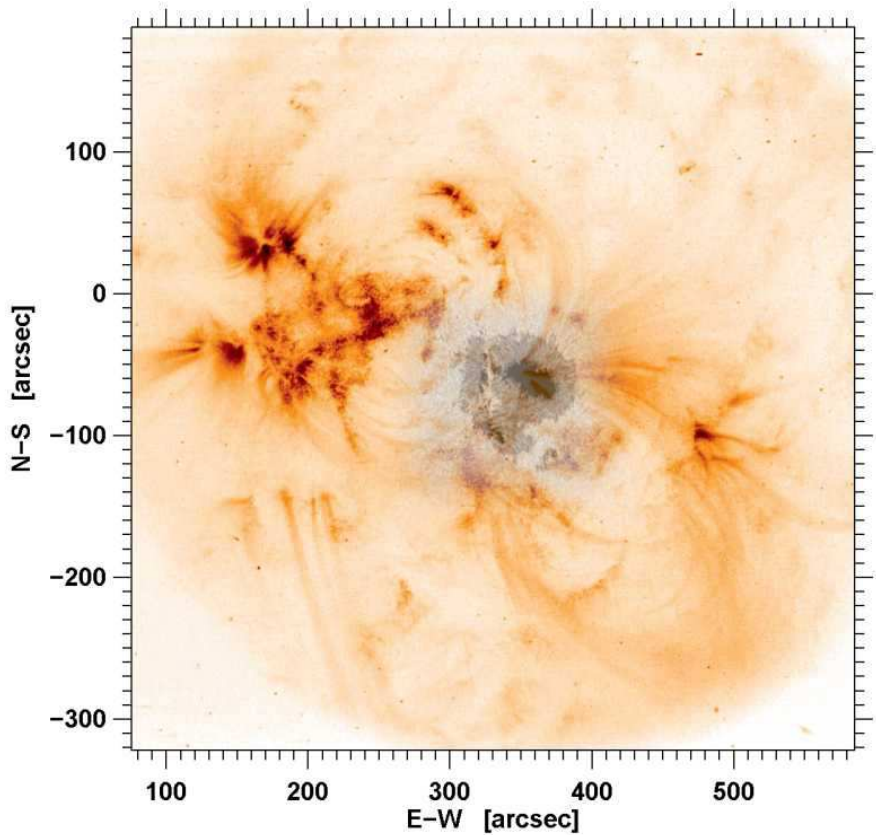}}
\caption{Superposition of a gray-scale Ca\,{\sc i} 610.3~nm line-wing
    filtergram observed at 17:52~UT on 2~May 2005 and a co-aligned, false-color,
    negative 171~\AA\ image obtained with TRACE at 03:48~UT on 3~May 2005
    showing the coronal morphology of active region NOAA~10756 at about
    $1 \times 10^6$~K.}
\label{FIG10}
\end{figure}

We used data of the {\it Transition Region and Coronal Explorer} (TRACE,
\opencite{Handy1999}) to compare the chromospheric with the coronal loop
structure above NOAA~10756. Figure~\ref{FIG10} is a superposition of a Ca\,{\sc
i} 610.3~nm line-wing filtergram and a negative TRACE 171~\AA\ image, which
maps coronal loop structures with temperatures of about $1 \times 10^{6}$~K.
Note that TRACE observations of NOAA~10756 were not available during the BBSO
observations. Therefore, we selected the first available image  on the
following day. In general, the chromospheric and the coronal loop topology is
very similar. Again, the absence of loop structures (Box~1 in
Figure~\ref{FIG09}) above the sunspots is worth noticing. As expected, the
low-lying loop system related to the sunspot moat is invisible in the TRACE
171~\AA\ image. However, the other two loop systems are clearly visible. The
loop system in the North-East, which is associated with the extensive plage
region, is routed in the linear group of sunspots ({\it N}$_2$) and the main
sunspot ({\it N}$_1$). On the other side of NOAA~10756, the loop system related
to the $\delta$ configuration ({\it P}$_1$) is much more pronounced compared to
the chromospheric observation. Several individual loops are contained in this
loop system.

%
%

\section{Discussion}

The most striking features in LCT and LOS velocity maps are localized areas of
strong flows. They fall into two categories: either they are associated with
small sunspots and pores embedded in a triangular region of granulation ({\it
G}$_2$), which is enclosed in the sunspot group, or they are located on
opposite sides of the magnetic polarity inversion line in regions of strong
shear flows (diagonal box {\it PU} in Figure~\ref{FIG02}). In the first case,
the observed downflows are typical for small sunspot and pores. Typically, the
velocity field of pores is close to zero near their center with strong
downflows in their periphery ({\it e.g.}, \opencite{Leka1998};
\opencite{Sankarasubramanian2003}).

The case of the flow kernels near the magnetic neutral line is different. These
kernels are not associated with dark features but with the sheared penumbral
velocity field. \inlinecite{MartinezPillet1994} reported strong, isolated
downflows near the neutral line of a $\delta$ spot, which were as large as
14~km~s$^{-1}$. The downward motions were accompanied by upward flows of about
1.5~km~s$^{-1}$ in a nearby region. These velocities were measured with the
{\it Advanced Stokes Polarimeter} (ASP) at NSO/Sacramento Peak by determining
the zero-crossing of Stokes $V$ profiles. This velocity signal might not be as
easily visible in the Doppler shift of the Stokes $I$ signal. Since our
spectral resolution is inferior to the spectrograph-type ASP instrument and we
have to rely on Stokes $I$ line profiles, the velocity signal of the flow
kernels in our study is not expected to be as high. The strongest upflow was
encountered in the disk-side penumbra of the main sunspot, which is accompanied
by kernel with smaller downflows on the other side of the polarity inversion
line. Despite the discrepancy in the velocity signal both observations of flow
kernels show a variety of commonalities. The flow kernels appear as pairs of
opposite flows on both sides of the strongest magnetic field gradient between
the penumbrae of the $\delta$ spot and main spot. They are long-lived features
(30~minutes to 3~hours). Typical sizes are in the range from $2^{\prime\prime}$
to $5^{\prime\prime}$. According to \inlinecite{MartinezPillet1994}, non-linear
phenomena in the transition region and chromosphere are responsible for the
observed high speeds. The strongly-curved penumbral filament in the limb-side
penumbra of the $\delta$ spot, which are almost tangential to the umbra,
indicate the presence of strong magnetic twist. \inlinecite{Leka2005} showed
that the onset of kink instability requires sufficient magnetic twist (flux
tubes with more than $2\pi$ winds) in the solar active region. At this moment,
it is not clear if two-dimensional spectro-polarimetry can provide enough
information to quantify the amount of twist. However, it can identify emerging
helicity-carrying flux tubes and help to identify areas, which are susceptible
to kink-instability. These quantities are valuable for any operational space
weather monitoring and predicting system.

$\delta$ spots have been identified as the most flare-productive active regions
\cite{Zirin1987}. Why does then the $\delta$ configuration in NOAA~19756
produce so little activity? For one, the $\delta$ configuration is not a very
close one, {\it i.e.}, it is not an ``island configuration''. Indeed, one could
argue that the $\delta$ spot is not located in the penumbra of the main spot
but possesses its own penumbra, which is just in very close proximity to the
main spot's penumbra. The penumbrae seem to merge in low-resolution
observations. However, the high-resolution observations clearly reveal the
distinct identity of both the $\delta$ and main spot. This observation is
confirmed by the LOS and horizontal flow patterns associated with the two
penumbrae. \inlinecite{Sammis2000} have provided a comprehensive study of the
relationship between flares and $\delta$ spots. $\beta\gamma\delta$ regions are
the most prolific flaring regions producing many more large flares compared to
other regions of similar size. $\beta\delta$ spots, such as NOAA~10756, are
ranked (see Figure~2 in \opencite{Sammis2000}) just below $\beta\gamma\delta$
spots on a par with $\delta$ spots. We would expect predominantly M-class flares
with some C-class and occasional X-class flare from a $\beta\delta$ region.
Considering the size of NOAA~10756 (about 1000 millionths of a solar
hemisphere), the likelihood for producing M-class flares was given.

In the region of colliding penumbrae (diagonal box {\it PU} in
Figure~\ref{FIG02}), we observe a negative (left-handed) magnetic shear along
the neutral line (Figure~\ref{FIG07}). The observed photospheric shear flow,
which is carried by the more horizontal Evershed flow channels, would tend to
decrease the magnetic shear, thus leading to a more potential configuration of
the local magnetic field. The temporal evolution of the magnetic shear in
region {\it PU} shows a fairly constant negative shear angle of about
$-36^{\circ}$, whereas in other regions of the neutral line the magnetic shear
increases (see Figure~\ref{FIG08}). The gradual build-up of magnetic shear in
these regions can be attributed to motions of the small $\delta$ configuration
and main sunspot on larger scales. The lack of major flare activity could,
therefore, also have its explanation in the fact that locally, the photospheric
shear flows might inhibit an increase of magnetic shear. At least it remained
constant at a level, which is common for non-$\delta$ regions in sunspots
\cite{Hagyard1984}. This situation is very different from the instantaneous
magnetic shear evolution observed during major flares \cite{Wang1994} and
differs from recent observations of photospheric shear flows in flaring active
region NOAA~10486 (\opencite{Yang2004}; \opencite{Deng2006}).

Based on our observations, we conclude that the polarity inversion line was too
short and sufficient shear was not available to create a channel enabling the
formation of an active region filament. The chromospheric and coronal loops are
pointing in two opposite directions leaving the volume above the sunspot group
almost void of any loops. Therefore, the magnetic field topology is much
different from the ``spine'' fields and their associated ``fan'' surface
surrounding a parasitic polarity, which have been observed in many flares with
filament eruptions and accompanying CMEs, {\it e.g.},
\inlinecite{Aulanier2000}, who studied the ``Bastille Day'' flare on 1998
July~14.

%
%

\section{Conclusions}

We have presented the first science observations with newly developed
visible-light post-focus instruments at BBSO. The combination of a high-order AO
system with an imaging spectrometer and a fast CCD camera system for speckle
observations enabled us to measure the three-dimensional flow fields in and
around solar active region NOAA~10756.

We found well-defined flow kernels embedded in a $\delta$ configuration, where
both the horizontal and LOS velocities reached up to 1.8~km~s$^{-1}$ and 2.5~km
s$^{-1}$, respectively. These flow kernels, which are one signature of the
atypical flow structure in $\delta$ spots, were previously associated with
flares ({\it e.g.}, \opencite{Yang2004}; \opencite{Deng2006}). However, since
these flow kernels are present in this relatively quiet active region, we
conclude that they are common features in magnetically complex active regions.
The observed photospheric flows are related to the magnetic fields of the more
horizontal Evershed flow channels. A connection to the more vertical fields of
the uncombed penumbra still needs to be established, since these are the fields
straddling the magnetic neutral line. However, this task requires future
observations with diffraction-limited spectro-polarimetry with good spectral
resolution. However, the well-established flow interface between the opposite
polarity penumbrae in NOAA~10756 did not have a counterpart in the respective
magnetic flux systems. Here, the situation is more complex. Small-scale
intrusions of opposite polarity flux into the $\delta$ spot penumbra might
locally increase the flow speed, while on the other side, locally contribute to
the dissipation of magnetic flux. We find that the magnetic shear remains
fairly constant in the regions with the photospheric shear flows and that the
direction of the flow might locally decrease magnetic shear. This scenario
hints at an intricate interaction between the global flow field established by
two deflected penumbrae and localized phenomena in penumbrae such as flow
kernels and inclusions of opposite polarity features. In any case, the peculiar
flows in the vicinity of $\delta$ spots are indicators for rapid field
evolution, flux emergence, and the presence of twisted fields.

Comparing our high-resolution observations with active region NOAA~10486
(\opencite{Xu2004b}; \opencite{Deng2006}), we find that the polarity inversion
line separating the $\delta$ spot from the main spot is much shorter. Only a
single location of strong shear flows was identified. Two intrusions of
negative polarity flux into areas with opposite-polarity plage limit the region
along the polarity inversion line where shear could be built-up. Thus, the
short magnetic neutral line is not conducive to the formation of an active
region filament. In addition, the overlying coronal and chromospheric field
topology as deduced from TRACE 171~\AA\ and BBSO H$\alpha$ filtergrams is much
different from the non-potential field above a filament channel, {\it i.e.} two
loop systems are deflected in opposite direction pointing away from the central
part of NOAA~10756. Therefore, magnetic reconnection above the polarity
inversion line is highly unlikely.

In summary, observations with high temporal, spatial, and spectral resolution
as well as with sufficient magnetic sensitivity are vital in advancing our
understanding of solar eruptive phenomena and how they affect Earth and the
near-Earth environment.

%
%

\acknowledgements We would like to thank Wolfgang Schmidt and an anonymous
referee, whose careful reading of the manuscript led to substantial
improvements. This work was supported by NSF under grant ATM 00-86999, ATM
02-36945, IIS~ITR 03-24816 and AST~MRI 00-79482 and by NASA under grant NAG
5-12782. The National Solar Observatory is operated by the Association of
Universities for Research in Astronomy under a cooperative agreement with the
National Science Foundation, for the benefit of the astronomical community. The
data analysis including the graphics was performed in the {\it Interactive Data
Language} (IDL) by ITT Visual Information Solutions.

%
%

\bibliographystyle{spr-mp-sola}
\bibliography{spr-mp-sola-jour,cdenker}

\end{article}
\end{document}